\documentstyle[aps]{revtex}

\draft

\begin{document}
\title{Separability and Fourier representations of density matrices}
\author{Arthur O. Pittenger \dag \footnote[3]{Present address: The Centre 
for Quantum Computation, Clarendon Laboratory,
 Oxford University}  and Morton H. Rubin \ddag}
\address{\dag Department of Mathematics and Statistics,
University of Maryland, Baltimore County,
Baltimore, MD 21228-5398}
\address{\ddag Department of Physics,
University of Maryland, Baltimore County,
Baltimore, MD 21228-5398}
\date{January 5, 2000}
\maketitle

\begin{abstract}
Using the finite Fourier transform, we introduce a generalization of
Pauli-spin matrices for $d$-dimensional spaces, and the resulting set of
unitary matrices $S\left( d\right) $ is a basis for $d\times d$ matrices. If 
$N=d_{1}\times d_{2}\times\cdots\times d_{b}$ and $H^{\left[ N\right] }=\bigotimes H^{%
\left[ d_{k}\right] }$, we give a sufficient condition for separability of a
density matrix $\rho $ relative to the $H^{\left[ d_{k}\right] }$ in terms
of the $L_{1}$ norm of the spin coefficients of $\rho .$ Since the spin
representation depends on the form of the tensor product, the theory applies
to both full and partial separability on a given space $H^{\left[ N\right] }$%
. It follows from this result that for a prescribed form of separability,
there is always a neighborhood of the normalized identity in which every
density matrix is separable. We also show that for every prime $p$ and $n>1$
the generalized Werner density matrix $W^{\left[ p^{n}\right] }\left(
s\right) $ is fully separable if and only if $s\leq \left( 1+p^{n-1}\right)
^{-1}$.
\end{abstract}


\section{Introduction}

One of the predictions of quantum mechanics is that spatially separated
components of a system can be entangled. The consequent prediction of
non-classical correlations among the separated components of a quantum
system has led to critiques of the foundations of quantum mechanics, as in
the famous Einstein, Podolsky, Rosen paper \cite{epr}, and to experiments
that have confirmed the predicted non-classical correlations, as in \cite
{asp}. Interest in entangled systems has been heightened by proposed
applications in quantum computation, for example \cite{shor}, and in quantum
communication, as exemplified most dramatically by teleportation \cite{benbr}%
. As a result there have been many publications which have examined various
aspects of entanglement, its measurement, and its use in quantum
communication such as references \cite{bennt,diV,niel,vidal,dur} to mention
only a few recent papers.

In this paper we shall be interested in the separability properties of
quantum systems in states defined on finite dimensional Hilbert spaces $%
H=H_{1}\otimes \cdots \otimes H_{n}$, where the $H_{k}$ denote the Hilbert
spaces of the subsystems. A state specified by a density matrix $\rho $ is
said to be completely separable on $H$ if it is a convex combination of
tensor products: 
\begin{equation}
\rho =\sum_{a}p\left( a\right) \rho ^{\left( 1\right) }\left( a\right)
\otimes \cdots \otimes \rho ^{\left( n\right) }\left( a\right) ,
\label{totsep}
\end{equation}
where $\rho ^{\left( k\right) }\left( a\right) $ is a density matrix on $%
H_{k}$. Since the same $\rho $ can have different convex representations, it
has proven difficult to determine generally applicable operational
conditions for separability, and determining such conditions is one of the
motivations for this paper. It is also possible to have different types of
separability by allowing sets of the subsystems to be entangled, cf. \cite
{dur,dur2}, and one can describe a lattice of levels of separability. The
theory we develop here applies to all of these various definitions of
separability.

A necessary condition for separability is that the partial transpose $\rho
^{T_{r}}$\ of a state $\rho $ should be a state \cite{peres}. If we
represent $\rho $\ as a matrix, this means that if $\rho =\left( \rho
_{j_{1}\ldots j_{n},k_{1}\ldots k_{n},}\right) $ then (taking $r=1$) 
\begin{equation}
\rho ^{T_{1}}=\left( \rho _{_{k_{1}j_{2}\ldots j_{n},j_{1}k_{2}\ldots
k_{n},}}\right)  \label{partrans}
\end{equation}
is also a density matrix. It is easy to confirm that if\ a density matrix is
separable, its partial transposes are also separable, but it has been shown 
\cite{hor3} that the converse is true only in the $2\otimes 2$ and $2\otimes
3$ cases. In the proof of this last result \cite{hor3}, a necessary and
sufficient criterion for separability was established, but there seems to be
no operational way of using this criterion as a general tool. Other studies
of separability, such as those in \cite{dur,schack,braun,caves} have found
operationally useful necessary conditions and sufficient conditions for
classes of densities or for special cases, but no general sufficient
conditions with a breadth of applicability analogous to that of the Peres
condition.

Broadly speaking, necessary conditions tend to be described in the
computational basis while sufficient conditions for $2$-level systems tend
to be described in terms of the Pauli spin basis. That observation motivated
the derivation of a change-of-basis formula in \cite{pitrub} which
facilitates the strategy of checking whether necessary conditions derived in
the computational basis are sufficient by using the (real) Pauli spin basis.
This approach leads to general sufficient conditions for full separability
which essentially give the condition in \cite{braun} as a corollary and also
leads to necessary and sufficient conditions for full separability of a
parametrized family of $n$-qubit densities which all satisfy the Peres
condition. The difficulty with extending this approach to $d$-level systems
is that the generally accepted definition of spin matrices as generators
of rotations does not capture the computationally useful features of the
Pauli matrices when $d\geq 3.$ One of the basic purposes of this paper is to
propose a general definition of $d$-level spin matrices which possess many
of those computational properties.

The Pauli matrices are special in that they are both Hermitian and unitary,
and together with the identity matrix $\sigma _{0}$ they form a basis of the
set of $2\times 2$ matrices. Our strategy is to generalize the role of the
Pauli matrices as a basis of unitary matrices at the expense of
Hermiticity. We show the applicability of these proposed $d$-level spin
matrices and in the appendix examine the $d=3$ case in some detail,
identifying properties analogous to those of the Pauli matrices. We also
define directly a general characterization of certain classes of trace one
projections of $d$--level systems. We use those projections to establish
necessary and sufficient conditions for full separability of generalized
Werner densities composed of any number $n$ of $d$--dimensional subsystems
for any prime $d$. In addition, we establish a general sufficient condition
for full or partial separability of densities of any dimension. Analogous
results for full separability were obtained in \cite{pitrub} for $d=2$ by
essentially the same methodology.

Other authors \cite{arv} have used a different set of operators in the $d=3$
case and some separability results were obtained recently in \cite{caves}.
Our proposed class is different, and we show that stronger separability
results can be obtained using these matrices and the\ strategy developed in 
\cite{pitrub}.

\section{A Necessary Condition}

As mentioned above, the Peres partial transpose condition is a general
necessary condition for separability \cite{peres}. In \cite{pitrub} a weaker
but useful condition was derived using the Cauchy-Schwarz inequality and has
the following application. Suppose $j=j_{1}\ldots j_{n}$ and $k=k_{1}\ldots
k_{n}$ differ in each component: $j_{r}\neq k_{r}$. Let $u$ and $v$ be
indices with $u_{r}\neq v_{r}$ and $\left\{ u_{r},v_{r}\right\} =$ $\left\{
j_{r},k_{r}\right\} .$ Then for fully separable\ states $\rho $ 
\begin{equation}
\left( \sqrt{\rho _{j,j}}\sqrt{\rho _{k,k}}\right) \geq \left| \rho
_{u,v}\right| ,  \label{necessary}
\end{equation}
where $\rho $ is written as a matrix in the computational basis defined by
the tensor products of $\left| j_{i}\right\rangle \left\langle k_{i}\right|
,1\leq i\leq n$. As an application, consider the following generalization of
the Werner density matrix \cite{werner} on the $N=d^{n}$ dimensional Hilbert
space $H^{\left[ N\right] }$: 
\[
W^{\left[ N\right] }\left( s\right) =\frac{1-s}{d^{n}}I+s\cdot \tau 
\]
where $I$ is the identity and $\tau $ is the projection defined by the state 
\begin{equation}
\left| \psi ^{\left[ N\right] }\right\rangle =\frac{1}{\sqrt{d}}\left(
\left| 0\ldots 0\right\rangle +\left| 1\ldots 1\right\rangle +\ldots +\left|
(d-1)\ldots (d-1)\right\rangle \right) .  \label{3nwerner}
\end{equation}
(In the sequel we let $\tilde{k}$ denote the repeated index $k\ldots k.$) In
the computational basis $W_{j,j}^{\left[ N\right] }\left( s\right) $ equals $%
\left( \frac{1-s}{d^{n}}+\frac{s}{d}\right) $ when $j$ is in $\left\{ \tilde{%
k}:0\leq k<d\right\} $ and equals $\frac{1-s}{d^{n}}$\ otherwise. The only
non-zero off-diagonal elements are $W_{\tilde{j},\tilde{k}}^{\left[ N\right]
}\left( s\right) =$ $\frac{s}{d}$. Choosing $j$ and $k$ appropriately in (%
\ref{necessary}), we have the necessary condition $1\geq s\left(
1+d^{n-1}\right) $. To show that this condition is also sufficient, we will
use the spin representation to prove $W^{\left[ N\right] }\left( s\right) $
is fully separable when $d$ is prime and $s=\left( 1+d^{n-1}\right) ^{-1}$.
In order to do that, however, we first need to define the spin
representation.\smallskip 

\section{Computational and Spin Bases}

Let $H^{\left[ N\right] }$ denote an $N$-dimensional Hilbert space where 
$N=d_{1}\times d_{2}\times\cdots\times d_{b}$. \ In this section we define different
bases for $N\times N$ matrices on $H^{\left[ N\right] }$ based on different
representations of $H^{\left[ N\right] }$ as a tensor product space, and the
discussion is purely mathematical. In the applications we will be concerned
with a specific representation $H^{\left[ N\right] }=\bigotimes_{a=1}^{b}H^{%
\left[ d_{a}\right] }$ and with the corresponding separability properties of
densities on $H^{\left[ N\right] }.$ The bases used will depend on the order
of the tensor product as will the representation of a density matrix as $%
\rho =\rho _{1}\otimes \cdots \otimes \rho _{b}$, the tensor product of
densities $\rho _{k}$ on the $H^{\left[ d_{k}\right] }$. For example, we
might want to examine separability of a density matrix on $H^{\left[ 90%
\right] }=H^{\left[ 6\right] }\otimes H^{\left[ 15\right] }$ using matrices
consistent with that tensor product. In a subsequent application one might
want $H^{\left[ 6\right] }$ to represent the tensor product of spin $1/2$
and spin $1$ particles, i.e. a tensor product of $H^{\left[ 2\right] }$ and $%
H^{\left[ 3\right] }$, and the order of the sub-tensor product shouldn't
affect the theory. We confirm that assertion by showing that permuting the
order of a tensor product corresponds to a conjugation operation and thus
that the theory is generally applicable with only notational changes for
particular applications.

\textbf{Lemma 1} Let $N=d_{1}\times d_{2}\times \cdots \times d_{b}$ and suppose $M$ is an $N\times N$
matrix with $M=C^{\left( 1\right) }\otimes \cdots \otimes C^{\left( b\right)
}$, where the $C^{\left( k\right) }$ are $d_{k}\times d_{k}$ matrices. Let $%
\sigma $ denote a permutation of $\left\{ 1,\ldots ,b\right\} $ and let $%
M_{\sigma }=C^{\left( \sigma \left( 1\right) \right) }\otimes \cdots \otimes
C^{\left( \sigma \left( b\right) \right) }$. Then for all such $C^{\left(
k\right) }$'s there is a permutation matrix $Q_{\sigma }$ such that $%
Q_{\sigma }M_{\sigma }Q_{\sigma }^{-1}=M$.

{\it Proof}: If $M\left( j,k\right) =C^{\left( 1\right) }\left(
j_{1},k_{1}\right) \cdot \cdots \cdot C^{\left( b\right) }\left(
j_{b},k_{b}\right) $, the index $j$ corresponds to the ordered $b$-tuple $%
\left( j_{1},\ldots ,j_{b}\right) $ and $j$ is uniquely defined by $%
j=j_{1}\left( d_{2}\times \cdots \times d_{b}\right) +j_{2}\left( 
d_{3}\times \cdots \times d_{b}\right) +\ldots +j_{b}$ and similarly for $k$. Let $%
j_{\sigma }$ correspond to $\left( j_{\sigma \left( 1\right) },\ldots
,j_{\sigma \left( b\right) }\right) $ and define the permutation matrix $%
Q_{\sigma }$ by $Q_{\sigma }\left( j,s\right) =\delta \left( j_{\sigma
},s\right) $. Then $Q_{\sigma }^{-1}=Q_{\sigma }^{t}$ and 
\[
Q_{\sigma }M_{\sigma }Q_{\sigma }^{-1}\left( j,k\right) =M_{\sigma }\left(
j_{\sigma },k_{\sigma }\right) =C^{\left( \sigma \left( 1\right) \right)
}\left( j_{\sigma \left( 1\right) },k_{\sigma \left( 1\right) }\right) \cdot
\cdots \cdot C^{\left( \sigma \left( b\right) \right) }\left( j_{\sigma
\left( b\right) },k_{\sigma \left( b\right) }\right) =M\left( j,k\right) ,
\]
completing the proof.

The motivation for this work comes from the identification of the $2\times 2$
Hadamard matrix as a key tool in working with $2$-level systems.
Specifically, suppose $\rho =\frac{1}{2}\left( \sigma _{0}+\sigma
_{m}\right) $, where $\sigma _{m}=\sum m_{j}\sigma _{j}$ and $\sigma
_{1}=\sigma _{x}$, $\sigma _{2}=\sigma _{y}$, and $\sigma _{3}=\sigma _{z}$
are the usual Pauli matrices. Then the coefficients in the spin basis are
related to those in the computational basis by the $2\times 2$ {\it Hadamard}
matrix: 
\begin{equation}
\left( 
\begin{array}{cc}
1 & m_{1} \\ 
m_{3} & -im_{2}
\end{array}
\right) =\left( 
\begin{array}{cc}
1 & 1 \\ 
1 & -1
\end{array}
\right) \left( 
\begin{array}{cc}
\rho _{00} & \rho _{01} \\ 
\rho _{11} & \rho _{10}
\end{array}
\right) .  \label{2coef}
\end{equation}
The matrices in the two bases are connected in a similar fashion: 
\begin{equation}
\left( 
\begin{array}{cc}
\sigma _{0} & \sigma _{1} \\ 
\sigma _{3} & i\sigma _{2}
\end{array}
\right) =\left( 
\begin{array}{cc}
1 & 1 \\ 
1 & -1
\end{array}
\right) \left( 
\begin{array}{cc}
E_{0,0} & E_{0,1} \\ 
E_{1,1} & E_{1,0}
\end{array}
\right) \text{,}  \label{2matr}
\end{equation}
where the matrices $E_{j,k}=|j\rangle \langle k|$ define the {\it %
computational} basis. Note that a systematic application of these
relationships requires both the use of the real Pauli matrices and a
reindexing of both the Pauli matrices and the computational basis matrices
to conform to the matrix notation.

The Hadamard matrix is the $2\times 2$ Fourier transform, and we extend the
idea of\ (\ref{2matr}) to $d\times d$ matrices. The {\it adjusted} basis $%
A=\left\{ A_{j,k},0\leq j,k<d\right\} $ is the set of $d\times d$ matrices
defined by $A_{j,k}=E_{j,j+k}$, where $+$ denotes addition modulo $d$, and
we {\it define} the ``spin'' matrices $S=\left\{ S_{j,k},\text{ }0\leq
j,k<d\right\} $ using the analogue of (\ref{2matr}) and the finite Fourier
transform. Thus $\left( S\right) \equiv F\cdot \left( A\right) $ where $%
F\left( j,k\right) =\exp \left( 2\pi ijk/d\right) =\eta ^{jk}$ with $\eta
=\exp \left( 2\pi i/d\right) $. (We will make the dependence on $d$ explicit
below.) In detail 
\begin{equation}
S_{j,k}=\sum_{r=0}^{d-1}F\left( j,r\right) A_{r,k}  \label{dmatrjk}
\end{equation}
is a sum of products of scalars times matrices. Since $F$ is invertible, it
follows that $S$ is also a basis for the $d\times d$ matrices. Note that (%
\ref{2matr}) is a special case of (\ref{dmatrjk}) with $d=2$ and $\eta =-1$.

To illustrate these ideas, it is useful to write out the results for $d=3$
in detail. Then $\eta =\exp \left( 2\pi i/3\right) $ and 
\begin{eqnarray*}
S_{00}=\left( 
\begin{array}{ccc}
1 & 0 & 0 \\ 
0 & 1 & 0 \\ 
0 & 0 & 1
\end{array}
\right) &S_{01}=\left( 
\begin{array}{ccc}
0 & 1 & 0 \\ 
0 & 0 & 1 \\ 
1 & 0 & 0
\end{array}
\right) &S_{02}=\left( 
\begin{array}{ccc}
0 & 0 & 1 \\ 
1 & 0 & 0 \\ 
0 & 1 & 0
\end{array}
\right) \\
S_{10}=\left( 
\begin{array}{ccc}
1 & 0 & 0 \\ 
0 & \eta & 0 \\ 
0 & 0 & \eta ^{2}
\end{array}
\right) &S_{11}=\left( 
\begin{array}{ccc}
0 & 1 & 0 \\ 
0 & 0 & \eta \\ 
\eta ^{2} & 0 & 0
\end{array}
\right) &S_{12}=\left( 
\begin{array}{ccc}
0 & 0 & 1 \\ 
\eta & 0 & 0 \\ 
0 & \eta ^{2} & 0
\end{array}
\right) \\
S_{20}=\left( 
\begin{array}{ccc}
1 & 0 & 0 \\ 
0 & \eta ^{2} & 0 \\ 
0 & 0 & \eta
\end{array}
\right) &S_{21}=\left( 
\begin{array}{ccc}
0 & 1 & 0 \\ 
0 & 0 & \eta ^{2} \\ 
\eta & 0 & 0
\end{array}
\right) &S_{22}=\left( 
\begin{array}{ccc}
0 & 0 & 1 \\ 
\eta ^{2} & 0 & 0 \\ 
0 & \eta & 0
\end{array}
\right)
\end{eqnarray*}

The spin matrices $S$ not only form a basis for $d\times d$ matrices{\bf ,}
but share many other properties with the real Pauli matrices, which we
record next. We should note that the matrices $S_{j,k}$ were also defined in
an earlier work by Fivel \cite{fivel} on Hamiltonians on discrete spaces,
and many of the properties listed below were first established there.

\textbf{Proposition 1}
Fix $d\geq 2$ and let $S$ denote the corresponding set of spin matrices. 
\newline
$(i)$ $S$ is an orthogonal basis of unitary matrices with respect to the
trace inner product. \newline
$(ii)$ If $d$ is odd, each matrix in $S$ is in $SU(d)$, while if $d$ is
even, $S_{j,k}$ is in $SU(d)$ if and only if $j+k$ is even. \newline
$(iii)$ $S_{j,k}=\left( S_{1,0}\right) ^{j}\cdot
\left( S_{0,1}\right) ^{k}$, $\left( S_{j,k}\right) ^{\dagger }=\eta
^{jk}S_{d-j,d-k}=(S_{j,k})^{d-1}$ and $\left[ S_{j,k},S_{r,s}\right] =\left(
\eta ^{kr}-\eta ^{js}\right) S_{j+r,k+s}$ using addition $\bmod \, d.$
\newline
$(iv)$ $tr\left( S_{j,k}\right) =0$ for all $\left( j,k\right) \neq \left(
0,0\right) .$

{\it Proof}: The key observation, noted in \cite{fivel}, is that the
matrices are generated by $S_{1,0}$ and $S_{0,1}$:$\ S_{j,k}=\left(
S_{1,0}\right) ^{j}\cdot \left( S_{0,1}\right) ^{k}$ with $S_{0,1}\cdot
S_{1,0}=\eta S_{1,1}$. All of the remaining assertions, including
orthogonality, follow from those relations and from easy computations. A
useful consequence of the manipulations is 
\begin{equation}
(S_{j,k})^{m}=\eta ^{(j\cdot k)m(m-1)/2}S_{mj,mk}\text{.}  \label{powerd}
\end{equation}

Unlike the Pauli matrices, these spin matrices need not be Hermitian; for
example{\bf ,} when $d=3$ only the identity matrix is Hermitian. Thus, when
computing the coefficients of a density matrix in these bases, as we do
next, the Hermitian conjugation notation has to be retained. Note that the
very last assertion in Corollary 1 corresponds to the usual inequality
relating the $L_{2}$ magnitude of a Fourier transform and the $L_{2}$
magnitiude of the original function.

\textbf{Corollary 1} 
$(i)$ The matrix elements of a $d\times d$ density matrix $\rho $ in the
different bases are related by $\left( s\right) =F^{\ast }\cdot \left(
a\right) $.\\
$(ii)$ $s_{0,0}=1$, $s_{d-j,d-k}=\eta ^{jk}s_{j,k}^{\ast }$ and 
$\frac{1}{d}(F\cdot s)_{j,0}=\rho _{j,\,j}\geq 0$. \\
$\left(iii\right)$
$\sum_{j,k}\left| s_{j,k}\right| ^{2}=d\sum \left| \rho _{j,k}\right| ^{2}$
and $\sqrt{\sum_{j,k}\left| s_{j,k}\right| ^{2}}\sqrt{\sum_{j,k}\left| \rho
_{j,k}\right| ^{2}}\geq 1/\sqrt{d}.$

{\it Proof}{\bf :} We expand an arbitrary density matrix in the two bases: 
\[
\rho =\sum_{j,k}a_{j,k}A_{j,k}=\frac{1}{d}\sum_{j,k}s_{j,k}S_{j,k}
\]
where $a_{j,k}=Tr\left( A_{j,k}^{\dagger }\rho \right) $ gives $a_{j,k}=\rho
_{j,\,j+k}$, using addition $\bmod \, d,$
and $s_{j,k}=Tr\left( S_{j,k}^{\dagger }\rho \right) $. Then from (\ref
{dmatrjk}), $s_{j,k}=\sum_{r=0}^{d-1}F^{\ast }\left( j,r\right) a_{r,k}$,
which proves $\left( i\right) $. Note that we have to include a complex
conjugation in the formula which is unnecessary in the $d=2$ case, since the
Hadamard matrix has real entries. The assertions in $\left( ii\right) $
follow from the definitions and from the fact that $\rho $ is Hermitian with
trace equal to one. Finally the relations in $\left( iii\right) $ follow
from $Tr\left( \left( s\right) ^{\dagger }\left( s\right) \right) =Tr\left(
\left( a\right) ^{\dagger }\left( F^{\ast }\right) ^{\dagger }\left( F^{\ast
}\right) \left( a\right) \right) =dTr\left( \left( a\right) ^{\dagger
}\left( a\right) \right) =dTr\left( \rho ^{2}\right) $, and from $Tr\left(
\rho ^{2}\right) =\sum_{k}\lambda _{k}^{2}\geq \sum_{k}1/d^{2}=1/d$, where
the $\lambda _{k}$ are the non-negative eigenvalues of the density $\rho.$ 

Now let $N=d_{1}\times d_{2}\times\cdots\times d_{b}$ with $d_{i}\geq 2$ and with the
order of multiplication fixed throughout the discussion. We use the
underlying and fixed tensor product representation of $H^{\left[ N\right] }$
to define the sets of computational and adjusted bases $E^{\left[ N\right] }$
and $A^{\left[ N\right] }$ for $N\times N$ matrices as 
\[
E_{j,k}^{\left[ N\right] }=E_{j_{1},k_{1}}^{\left( 1\right) }\otimes \cdots
\otimes E_{j_{b},k_{b}}^{\left( b\right) }\hspace{0.25in}\text{and\hspace{%
0.25in}}A_{j,k}^{\left[ N\right] }=A_{j_{1},k_{1}}^{\left( 1\right) }\otimes
\cdots \otimes A_{j_{b},k_{b}}^{\left( b\right) },
\]
where $j$ and $k$ correspond to their $b$-tuples and the superscripts in
parentheses identify the corresponding $d_{i}$. It follows that $A_{j,k}^{%
\left[ N\right] }=E_{j,j\oplus k}^{\left[ N\right] }$ where the addition of
the indices is defined by 
\begin{equation}
j\oplus k\equiv ( j_{1}+k_{1} \bmod \,d_{1},
\ldots ,j_{b}+k_{b} \bmod \, d_{b}) .  \label{modadd}
\end{equation}
The corresponding set of spin matrices $S^{\left[ N\right] }$ is then
defined by $\left( S^{\left[ N\right] }\right) =F^{\left[ N\right] }\left(
A^{\left[ N\right] }\right) $ or
\[
S_{j,k}^{\left[ N\right] }=\sum_{r=0}^{N-1}F^{\left[ N\right] }\left(
j,r\right) A_{r,k}^{\left[ N\right] }
\]
where $F^{\left[ N\right] }=F^{\left( 1\right) }\otimes \cdots \otimes
F^{\left( b\right) }$ is the usual tensor product of the Fourier transforms $%
F^{\left( k\right) }$ which depend on $d_{k}$. Since we will be taking
powers of the $\eta $'s, we will use subscripts to denote the dependency of $%
\eta $ on $d_{k}$: $\eta _{k}=\exp \left( 2\pi i/d_{k}\right) $. It is easy
to show that an equivalent definition of $S^{\left[ N\right] }$ is given by 
\begin{equation}
S_{j,k}^{\left[ N\right] }=\bigotimes_{i=1}^{b}\left( F^{\left( i\right)
}A^{\left( i\right) }\right) _{j_{i},k_{i}}\text{.}  \label{spindef}
\end{equation}

Linearity again implies that if $\rho ^{\left[ N\right] }$ is a density
matrix on the $N\times N$ Hilbert space $H^{\left[ N\right] }$ with 
\[
\rho ^{\left[ N\right] }=\sum_{j,k}a_{j,k}^{\left[ N\right] }A_{j,k}^{\left[
N\right] }=\frac{1}{N}\sum_{j,k}s_{j,k}^{\left[ N\right] }S_{j,k}^{\left[ N%
\right] }, 
\]
then 
\begin{equation}
\left( s^{\left[ N\right] }\right) =F^{\ast \left[ N\right] }\cdot \left( a^{%
\left[ N\right] }\right)  \label{ncoef}
\end{equation}
and $a_{j,k}^{\left[ N\right] }=\rho _{j.j\oplus k}^{\left[ N\right] }$%
.\smallskip \smallskip\ Thus we have two different representations for a
density matrix $\rho ^{\left[ N\right] }$, and both of them depend on the
underlying tensor product representation of $H^{\left[ N\right] }$.

\section{ Sigma Variations}

A fully separable density matrix can be represented as a convex combination
of tensor products of pure states or trace one projections, and we need to
represent such $d\times d$ projections in a systematic fashion in the spin
basis. (All projections in this paper are trace one projections.) In the
Appendix we show how all trace one projections for $d=3$ can be represented
in a form completely analogous to the $d=2$ case, but for our immediate
purposes we only need to characterize a subclass. The motivation is given by
writing the particular $d=2$ projections $\frac{1}{2}\left( \sigma _{0}\pm
\sigma _{k}\right) $ as $P_{k}(r)=\sum_{m=0}^{1}\left( \left( -1\right)
^{r}\sigma _{k}\right) ^{m}$, where $r=0$ or $r=1$. Then $P_{k}(r)$ is the
average of the cyclic subgroup generated by $(-1)^{r}\sigma _{k}$, and since 
$\left( (-1)^{r}\sigma _{k}\right) ^{2}=\sigma _{0}$, the key property $%
P_{k}(r)\cdot P_{k}(r)=P_{k}(r)$ reduces to an exercise in group theory. The
generalization of this idea to arbitrary $d$ is immediate, and we first
treat the case when $d$ is prime.

\textbf{Proposition 2} 
Let $d=p\geq 2$ be prime. Let $u=(j,k)\neq (0,0)$ denote the index of a spin
matrix $S_{u}$, and let $r$ be an integer. Then if $p>2,$ the matrix 
\begin{equation}
P_{u}(r)\equiv \frac{1}{p}\sum_{m=0}^{p-1}\left( \eta ^{r}S_{u}\right) ^{m}
\label{projection}
\end{equation}
is a projection with unit trace. The assertion is also valid for $p=2$
provided  $iS_{u}$ is used in lieu of $S_{u}$ throughout when $u=(1,1).$

{\it Proof}: A matrix $P$ is a pure state or a trace one projection if it is
Hermitian, has trace $1,$ and $P^{2}=P$. First, $\left( \eta
^{r}S_{u}\right) ^{m}$ is proportional to $S_{mj,mk}$; consequently, 
it cannot be proportional to $S_{0,0}$ for $0<m<p$. Therefore,
only $S_{0,0}$ contributes to the trace of $P\left( u,r\right) $, confirming
the trace condition. Using (\ref{powerd}) it follows that $\left( \left(
\eta ^{r}S_{u}\right) ^{m}\right) ^{\dagger }=\left( \eta ^{r}S_{u}\right)
^{p-m}$ and that when $p$ is odd $\left( \eta ^{r}S_{u}\right) ^{m}\left(
\eta ^{r}S_{u}\right) ^{p-m}=\left( S_{u}\right) ^{p}=\left( \eta
^{jk}\right) ^{\frac{p\left( p-1\right) }{2}}S_{0,0}=S_{0,0}$. Thus $P\left(
u,r\right) $ is Hermitian. The verification that $P\left( u,r\right)
^{2}=P\left( u,r\right) $ follows from an easy computation. The
assertion that $\left( \eta ^{jk}\right) ^{\frac{p\left( p-1\right) }{2}}=1$
fails for prime $p$ only when $p=2$ and $j=k=1$. Thus, the reintroduction of 
$i$ and of $-\sigma _{y}=iS_{1,1}$ is required to complete
the proof.

As an example of the notation, it is easy to check that if $k=0$, then $%
P_{j,0}(r) $ is one of the diagonal projections $E_{i,i}$. Other projections
are less sparse, however. For example, when $d=3$ and $k\neq 0$, $P_{j,k}(r) 
$ has no zero entries in the computational basis representation.

In the preceding proof, we exploited the fact that for $d$ an odd prime the
powers of each matrix $S_{u},u\neq (0,0),$ form a cyclic subgroup of order $%
d. $ When $d$ is not prime we can get analogous results using a similar
proof, but there are restrictions on the indices that arise since the
coefficient of the identity matrix in (\ref{powerd}) when $m=d$ need not be
unity. In Proposition 2 this led to the introduction of the factor $i=\exp
\left( \pi i/2\right) $ when $d=2$, and that modification is a special case
of a more general situation.

\textbf{Proposition 3} 
Suppose $d$ is composite. Let $u=\left( j,k\right) $ be $\left( 0,1\right) $%
, $\left( 1,0\right) $ or else an index such that $j\neq 0$ and $k\neq 0$
have no common factors. Suppose $d$ is odd or $j\cdot k$ is even. Then if $r$
is an integer, $P_{u}(r)=\frac{1}{d}\sum\limits _{m=0}^{d-1}(\eta^{r}S_{u})
^{m}$ is a projection with unit trace. If $d$ is even and $j\cdot k$ is odd,
then $P_{u}(r)=\frac{1}{d}\sum_{m=0}^{d-1}\left( \alpha \eta
^{r}S_{u}\right) ^{m}$ is a projection with unit trace, where $\alpha
=e^{\pi i/d}$.

{\it Proof}: Suppose $\left( \eta ^{r}S_{u}\right) ^{m}$ or $\left( \alpha
\eta ^{r}S_{u}\right) ^{m}$ is proportional to $S_{0,0}$ for $0<m<d$, so
that $mj=rd$ and $mk=sd$ for some integers $r$ and $s$. Since $j$ and $k$
are relatively prime, there are integers $a$ and $b$ such that $aj+bk=1$ 
\cite{Andrews}, and it follows that $m=ard+bsd=\left( ar+bs\right) d$,
contradicting $m<d$. Thus $P_{u}(r)$ has trace one. Using (\ref{powerd})
when $m=d$, we find that the coefficient of $S_{0,0}$ is one in the first
case, while in the second case the extra factor of $\alpha ^{d}=\left(
-1\right) $ is necessary to make the overall coefficient equal to one. In
both cases it follows from that key result as in Proposition 2 that $%
P_{u}^{2}(r)=P_{u}(r)$ and that $P_{u}(r)$ is Hermitian, completing the
proof.

An important relationship between these subgroup projections and the
generating spin matrix follows from the definitions.

\textbf{Corollary 2} 
For any integer $t\geq 0$ and any $d\geq 2$, 
\begin{equation}
(\eta ^{r}S_{u})^{t}=\sum_{m=0}^{d-1}\eta ^{-mt}P_{u}(m+r)\text{,}
\label{inversion}
\end{equation}
subject to the usual caveat about $\alpha $. In particular, $%
S_{0,0}=\sum_{m=0}^{d-1}P_{u}(m+r)$

{\it Proof}: $\sum_{m=0}^{d-1}\eta ^{-mt}P_{u}(m+r)=\frac{1}{d}%
\sum_{k}\left( \eta ^{r}S_{u}\right) ^{k}\sum_{m}\eta ^{-mt}\eta
^{mk}=\left( \eta ^{r}S_{u}\right) ^{t}$, as required.

Next consider a Hilbert space that is the direct product of $b$ Hilbert
spaces with dimensions $d_{1},\ldots ,d_{b}.$ Projections in the constituent 
$d_{i}$ dimensional spaces also define projections in tensor product spaces,
and the proof of the following is immediate. As before, we let the
superscript $k$ denote the dependence on $d_{k}$.

\textbf{Corollary 3} 
Let $N=d_{1}\times d_{2}\times \cdots\times d_{b}$ and let $H^{\left[ N\right]
}=\bigotimes_{a=1}^{b}H^{\left[ d_{a}\right] }$ be an $N$ dimensional
Hilbert space. Let $u$ denote a $b$-dimensional vector of index pairs $%
u_{i}=\left( j_{i},k_{i}\right) $ where $0\leq j_{i},k_{i}\leq d_{i}-1$, and
let $r=\left( r_{1},\ldots ,r_{b}\right) $ where the $r_{i}$ are integers.
Then if the $P_{u_{k}}^{\left( k\right) }(r_{k})$ are trace one projections
on $H^{[d_{k}]}$, 
\[
P_{u}(r)=\bigotimes_{k=1}^{b}P_{u_{k}}^{(k)}(r_{k})
\]
is a trace one projection on $H^{[N]}$, provided $\alpha S_{u}$ is used in
place of $S_{u}$ when $d$ is even and $u=(j,k)$ with $j\cdot k$ odd.
Furthermore, if $\eta \left( r\right) \equiv \prod_{k=1}^{b}\eta _{k}^{r_{k}}
$ and $t$ is a non-negative integer,
\begin{equation}
\left( \eta \left( r\right) S_{u}^{[N]}\right)
^{t}=\sum_{l_{1}=0}^{d_{1}-1}\ldots
\sum_{l_{b}=0}^{d_{b}-1}\bigotimes_{k=1}^{b}\eta
_{k}^{-l_{k}t}P_{u_{k}}^{\left( k\right) }(l_{k}+r_{k})\text{,}
\label{projsum}
\end{equation}
and in particular $S_{0,0}^{[N]}=\sum\limits_{l_{1}=0}^{d_{1}-1}\ldots
\sum\limits_{l_{b}=0}^{d_{b}-1}%
\bigotimes_{k=1}^{b}P_{u_{k}}^{(k)}(l_{k}+r_{k})$.

In order to show separability results for Werner densities, we need to
identify a special class of fully separable density matrices in the tensor
product space $H^{\left[ N\right] }\left( d\right) $ of $n$ $d$-dimensional
Hilbert spaces, where $N=d^{n}$. This approach is motivated by results in 
\cite{pitrub} and is our final variation on the Pauli $\sigma $-matrices.

\textbf{Proposition 4} 
Let $d\geq 2$ and let $u^{\left( n\right) }=\left( u_{1},\ldots
,u_{n}\right) $ and $r^{\left( n\right) }=\left( r_{1},\ldots ,r_{n}\right) $
denote vectors of indices and values as defined in the preceding
propositions. Then provided $\alpha S_{u}$ is used in place of $S_{u}$ when $%
d$ is even and $u=\left( j,k\right) $ with $j\cdot k$ odd, 
\[
\rho \left( u^{\left( n\right) },r^{\left( n\right) }\right) =\frac{1}{d^{n}}%
\left( \left( S_{0,0}\otimes \cdots \otimes S_{0,0}\right)
+\sum_{m=1}^{d-1}\left( \left( \eta ^{r_{1}}S_{u_{1}}\right) ^{m}\otimes
\cdots \otimes \left( \eta ^{r_{n}}S_{u_{n}}\right) ^{m}\right) \right) 
\]
is a fully separable density matrix on $H^{\left[ N\right] }\left( d\right) $%
.

{\it Proof}: The assertion is true for $n=1$ and suppose it holds for $n$.
Let $u^{(n+1)}$ and $r^{(n+1)}$ be given index and parameter vectors. Since
we require only the $n^{\prime }th$ and $(n+1)^{\prime }st$ indices in the
proof, we leave the other indices fixed and implicit and let $\rho \left(
u_{n},r_{n}\right) $ denote $\rho \left( u^{\left( n\right) },r^{\left(
n\right) }\right) $. By the induction hypothesis 
\[
\frac{1}{d}\sum_{s=0}^{d-1}\rho (u_{n},r_{n}+s)\otimes P_{u_{n+1}}^{\left(
n+1\right) }(r_{n+1}-s) 
\]
is fully separable. Multiplying out and collecting terms produces
expressions of the form 
\[
\lbrack (\eta ^{r_{1}}S_{u_{1}})^{m_{1}}\otimes \cdots \otimes (\eta
^{r_{n}}S_{u_{n}})^{m_{1}}]\otimes (\eta ^{r_{n+1}}S_{u_{n+1}})^{m_{2}}\frac{%
1}{d}\sum_{s=0}^{d-1}\eta ^{s(m_{1}-m_{2})}. 
\]
By the same analysis used earlier, terms with $m_{1}=m_{2}$ have an overall
coefficient of $1$ while all other terms have coefficient $0$, and that
completes the proof of the induction step.$\smallskip \smallskip \smallskip $

\section{Applications}

We now have the tools to prove a sufficient condition for full and partial
separability which extend the results in \cite{pitrub}. This is a general
sufficient condition for full or partial separability, and the results in 
\cite{braun} and \cite{caves} for $N=2^{n}$ and $N=3^{n}$ respectively
showing the existence of a neighborhood of the weighted identity in which
every density is fully separable follow as corollaries. As usual $H^{\left[ N%
\right] }$ will denote an $N$ dimensional Hilbert space which can be written
as a tensor product: $H^{\left[ N\right] }$ $=H^{\left[ d_{1}\right]
}\otimes \cdots \otimes H^{\left[ d_{b}\right] }$, where the $H^{\left[ d_{k}%
\right] }$ are $d_{k}$-dimensional spaces and 
$N=d_{1}\times d_{2}\times\cdots\times d_{b}$.
 We define $D\equiv \left( d_{1},\ldots ,d_{b}\right) $ and refer to $%
H^{\left[ d_{1}\right] }\otimes \cdots \otimes H^{\left[ d_{b}\right] }$ as
the $D$ tensor product version of $H^{\left[ N\right] }$. Since $H^{\left[ N%
\right] }$ may be represented as a tensor product space in different ways,
the kind of separability to be discussed depends on the representation. For
example, if $N=3^{n}$ and $H^{[N]}$ is represented as the tensor product of $%
n$ three-dimensional spaces, we are discussing full separability. If subsets
of the trits are taken together and represented in $3^{k}$-dimensional
spaces, we are discussing the corresponding partial separability. By virtue
of Lemma 1, we know that the fundamental mathematics involved doesn't depend
on the order in which the tensor products are taken or which trits are
grouped together.

In expressing the condition of the theorem, we use the $D$ spin coefficients
to introduce an $L_{1}$ norm on the space of $N\times N$ densities, and we
will refer to that hereafter as the $D$ spin norm and to the related
separability as $D$ separability.

\textbf{Theorem 1}
Let $H^{\left[ N\right] }$ denote an $N$-dimensional Hilbert space with $%
N=d_{1}\times d_{2} \times \cdots \times  d_{b}$. Suppose $H^{\left[ N\right] }=H^{\left[
d_{1}\right] }\otimes \cdots \otimes H^{\left[ d_{b}\right] }$, where the $%
H^{\left[ d_{k}\right] }$ are $d_{k}$-dimensional Hilbert spaces. If $\rho $
is a density matrix on $H^{\left[ N\right] }$, then $\rho $ is $D\equiv
\left( d_{1},\ldots ,d_{b}\right) $ separable provided 
\begin{equation}
\left\| \rho \right\| _{1,D}\equiv \sum_{\left( j,k\right) \neq \left(
0,0\right) }\left| s_{j,k}^{\left[ N\right] }\right| \leq 1,  \label{norm}
\end{equation}
where $\rho $ has the spin representation $\frac{1}{N}\sum_{j,k}s_{j,k}^{%
\left[ N\right] }S_{j,k}^{\left[ N\right] }$ defined in term of the $D$
tensor product: $S_{j,k}^{\left[ N\right] }=%
\bigotimes_{i=1}^{b}S_{j_{i},k_{i}}^{\left( i\right) }$. It follows that in
the set of density matrices on $H^{[N]}$ there is a neighborhood relative to 
$D$ of the random state $\frac{1}{N}S_{0,0}^{[N]}$ in which every density
matrix is $D$ separable.

{\it Proof}: If $d_{i}$ is prime or $j_{i}$ and $k_{i}$ are relatively
prime, the factor $S_{j_{i},k_{i}}^{\left( i\right) }$ can be written as a
weighted sum of projections as in Corollary 2. If $d_{i}$ is composite and
the indices $j_{i}$ and $k_{i}$ are not relatively prime, then up to a
factor of $\eta _{i}^{t_{i}}$, $S_{j_{i},k_{i}}^{\left( i\right) }$ can be
written as $\left( \eta _{i}^{r}S_{u}^{\left( i\right) }\right) ^{s}$ for
some $u=\left( \bar{j}_{i},\bar{k}_{i}\right) $ with $\bar{j}_{i}$\ and $%
\bar{k}_{i}$\ relatively prime, and thus $S_{j_{i},k_{i}}^{\left( i\right) }$
can also be written as a weighted sum of projections. Now since $\rho $ is a
density, either $S_{j,k}^{\left[ N\right] }$ is Hermitian and thus $s_{j,k}^{%
\left[ N\right] }$ is real, or $S_{j,k}^{\left[ N\right] }$ appears in a
pair $s_{j,k}^{\left[ N\right] }S_{j,k}^{\left[ N\right] }+s_{j,k}^{\ast %
\left[ N\right] }\left( S_{j,k}^{\left[ N\right] }\right) ^{\dagger }$. In
the second case we use (\ref{projsum}) in Corollary 3 and the preceding
comments to collect the various factors of $\eta _{i}^{t_{i}}$ together and
obtain 
\begin{eqnarray*}
s_{j,k}^{[N]}S_{j,k}^{[N]}+s_{j,k}^{\ast \lbrack N]}\left(
S_{j,k}^{[N]}\right) ^{\dagger }&=&\sum_{l_{1}=0}^{d_{1}-1}\ldots
\sum_{l_{b}=0}^{d_{b}-1}\bigotimes_{k=1}^{b}P_{u_{k}}^{(k)}(l_{k}+r_{k})%
\left\{ \beta _{j,k}s_{j,k}^{[N]}\eta ^{\ast }\left( l\right) +\beta
_{j,k}^{\ast }s_{j,k}^{\ast \lbrack N]}\eta \left( l\right) \right\}  \\
&=&\left| s_{j,k}^{\left[ N\right] }\right| \sum_{l_{1}=0}^{d_{1}-1}\ldots
\sum_{l_{b}=0}^{d_{b}-1}\bigotimes_{k=1}^{b}P_{u_{k}}^{(k)}(l_{k}+r_{k})%
\left\{ \exp \left( i\theta _{j,k}\right) \eta ^{\ast }\left( l\right) +\exp
\left( -i\theta _{j,k}\right) \eta \left( l\right) \right\} 
\end{eqnarray*}
where $\theta _{j,k}$ denotes the phase of $\beta _{j,k}s_{j,k}^{[N]}$ and $l
$ denotes the $b$-vector with components $l_{k}$. The caveat that $\alpha
_{i}S_{u}$ is in the projections $P_{u_{k}}^{(i)}(l_{k}+r_{k})$ in lieu of $%
S_{u}$ when $d_{i}$ is even and $u=\left( j_{i},k_{i}\right) $ with $%
j_{i}k_{i}$ odd applies throughout the proof and will not be explicitly
cited. Since $\alpha _{i}$ has magnitude $1$, only the phase factor will be
affected. Using the last assertion in Corollary 3, we can write $\left|
s_{j,k}^{\left[ N\right] }\right| S_{0,0}^{\left[ N\right] }+\frac{1}{2}%
\left( s_{j,k}^{\left[ N\right] }S_{j,k}^{\left[ N\right] }+s_{j,k}^{\ast %
\left[ N\right] }\left( S_{j,k}^{\left[ N\right] }\right) ^{\dagger }\right) 
$ as 
\[
\left| s_{j,k}^{\left[ N\right] }\right| \sum_{l_{1}=0}^{d_{1}-1}\ldots
\sum_{l_{b}=0}^{d_{b}-1}\bigotimes_{k=1}^{b}P_{u_{k}}(l_{k}+r_{k})\left\{
1+\cos \left( \theta _{j,k}-\arg \left( \eta \left( l\right) \right) \right)
\right\} .
\]
Since the expression in brackets is non-negative, the right-hand side is a
non-negative multiple of a $D$-separable density. In the case when $S_{j,k}^{%
\left[ N\right] }$ is Hermitian we derive the same expression with the same
conclusion. It follows that $\rho $ can be written as a convex combination
of fully separable densities plus the residual term $\left( 1-\sum_{\left(
j,k\right) \neq \left( 0,0\right) }\left| s_{j,k}^{\left[ N\right] }\right|
\right) \frac{1}{N}S_{0,0}^{\left[ N\right] }$. The hypothesis guarantees
that the coefficient of $\frac{1}{N}S_{0,0}^{\left[ N\right] }$\ is
non-negative, and that completes the proof of $D$-separability.\smallskip 

As another application of the machinery, we can prove for prime $p$ that the
necessary condition $s\leq \left( 1+p^{n-1}\right) ^{-1}$ is sufficient for
full separability of the generalized Werner density matrix $W^{\left[ N%
\right] }\left( s\right) =\frac{1-s}{N}$ $I+s\cdot \tau $. We have $N=p^{n}$%
, $I$ is the identity, $\tau $ is the projection defined by the state $%
\left| \psi ^{\left[ N\right] }\right\rangle =\frac{1}{\sqrt{p}}%
\sum_{k=0}^{p-1}\left| \tilde{k}\right\rangle $ and $\tilde{k}$ denotes the $%
n$-long repeated index $k\ldots k$. Given this special structure we find 
\[
W^{\left[ N\right] }\left( s\right) =\frac{1-s}{p^{n}}I+\frac{s}{p}%
\sum_{j=0}^{p-1}\sum_{k=0}^{p-1}\left| \tilde{j}\right\rangle \left\langle 
\tilde{k}\right| =\frac{1-s}{p^{n}}I+\frac{s}{p}\sum_{j=0}^{p-1}%
\sum_{k=0}^{p-1}A_{\tilde{j},\tilde{k}}^{\left[ N\right] } 
\]
where we have used the modular vector addition defined in (\ref{modadd}).
Computing the spin coefficients gives $s_{0,0}=1$, $s_{j,m}=0$ if $m$ is not
a $\tilde{k}$ with $0\leq k<p$, and otherwise 
\[
s_{j,\tilde{k}}=\sum_{r}F^{\ast }\left( j,r\right) \frac{s}{p}\delta \left(
r,Ind\right) 
\]
where $Ind=\left\{ \tilde{r}:0\leq r<p\right\} $. Using the dot product of
the index vectors $j\bullet r=\sum j_{k}r_{k} \bmod \, p,$
\[
s_{j,\tilde{k}}=\sum_{r}\exp \left( \frac{-2\pi i}{p}\left( j\bullet
r\right) \right) a_{r,\tilde{k}}=\frac{s}{p}\left( 1+\sum_{r=1}^{p-1}\exp
\left( \frac{-2\pi i}{p}\left( j\bullet \tilde{r}\right) \right) \right) . 
\]
Let $Ind\left( p,n\right) =\{ j:\sum_{r=0}^{N-1}j_{r}=0 \bmod \, p \}.$ 
Then it's easy to check that $s_{j,\tilde{k}}=s$ if and only if
\ $j$ is in $Ind\left( p,n\right) $ and that there are exactly $p^{n-1}$
such indices. All other $s_{j,\tilde{k}}$ equal zero, and we can write $W^{%
\left[ N\right] }\left( s\right) $ in the spin basis as

\begin{equation}
W^{\left[ N\right] }\left( s\right) =\frac{1-s}{p^{n}}S_{0,0}^{\left[ N%
\right] }+s\sum_{j\in Ind\left( p,n\right) }\sum_{k=0}^{p-1}S_{j,\tilde{k}}^{%
\left[ N\right] }.  \label{spinrep}
\end{equation}
\qquad

\textbf{Theorem 2}
Let $p$ be prime and $N=p^{n}$. Then the generalized Werner density matrix $%
W^{\left[ N\right] }\left( s\right) $ is fully separable on $H^{\left[ n%
\right] }\left( p\right) $ if and only if $s\leq \left( 1+p^{n-1}\right)
^{-1}$.

{\it Proof}: As shown above, necessarily $s\leq \left( 1+p^{n-1}\right)
^{-1} $. Checking the preceding derivation, note that 
\[
\frac{1}{p}\sum_{j=0}^{p-1}E_{\tilde{j},\tilde{j}}=\frac{1}{p}%
\sum_{j=0}^{p-1}A_{\tilde{j},\tilde{0}}^{\left[ N\right] }=\sum_{j\in
Ind\left( p,n\right) }S_{j,0}^{\left[ N\right] } 
\]
is a sum of fully separable projections. Taking $s=\left( 1+p^{n-1}\right)
^{-1}$ we can write $W^{\left[ n\right] }\left( s\right) $ as 
\[
W^{\left[ n\right] }\left( s\right) =\frac{1}{1+p^{n-1}}\left[ \frac{1}{p}%
\sum_{j=0}^{p-1}E_{\tilde{j},\tilde{j}}+\sum_{j\in Ind(p,n)}\frac{1}{p^{n}}%
\left( S_{\tilde{0},\tilde{0}}+\sum_{k=1}^{p-1}S_{j,\tilde{k}}\right) \right]
\text{.} 
\]
For each $k\neq 0$, $Ind(p,n)$ is mapped in a one-to-one manner onto itself
by $j\rightarrow kj$ where $\left( kj\right) _{r}=kj_{r} \bmod \, p.$
Thus \ 
\begin{equation}
W^{\left[ n\right] }\left( s\right) =\frac{1}{1+p^{n-1}}\left[ \frac{1}{p}%
\sum_{j=0}^{p-1}E_{\tilde{j},\tilde{j}}+\sum_{j\in Ind(p,n)}\left( \frac{1}{%
p^{n}}\left( S_{\tilde{0},\tilde{0}}+\sum_{k=1}^{p-1}S_{kj,\tilde{k}}\right)
\right) \right] \text{.}  \label{wersep}
\end{equation}
But since 
\[
\left( S_{j_{1},1}\right) ^{k}\otimes \cdots \otimes \left(
S_{j_{n},1}\right) ^{k}=\eta ^{k\sum j_{i}}S_{kj,\tilde{k}}=S_{kj,\tilde{k}} 
\]
for $j$ in $Ind(p,n)$, each $j$-sum in (\ref{wersep}) is fully separable by
Proposition 4, completing the proof.

It follows for the Werner densities that at the extreme value $s=\left(
1+p^{n-1}\right) ^{-1}$, $\sum_{\left( j,k\right) \neq \left( 0,0\right)
}\left| s_{j,k}^{\left[ N\right] }\right| =p\frac{1-p^{-n}}{1+p^{-\left(
n-1\right) }}$, where the coefficients are based on the decomposition $%
D=\left( p,\ldots ,p\right) .$ When $p=n=2$, that value is $1$, showing that
the global bound of Theorem 1 is attained.\ However, for larger $n$ and
prime $p\geq 2$ the condition $\left\| \rho \right\| _{1,D}\leq 1$ is too
strong for that class, and the special structure of the Werner densities
allowed a more refined analysis of $D=\left( p,\ldots ,p\right) $%
-separability.

It was shown in the qubit case in \cite{pitrub} that for each $n$ and given $%
\epsilon >0$, there exists a $D=\left( 2,\ldots ,2\right) $-inseparable
density on $H^{\left[ 2^{n}\right] }$ which has $\left\| \rho \right\|
_{1,D}<1+\epsilon $. Thus, for each fixed $n$\ the sufficient condition of
Theorem 1 is the best possible for full separability of qubits. We
conjecture that the same is true in general: given any separability vector $%
D $ and $\epsilon >0$ there exists a $D$-inseparable density $\rho $ with $%
\left\| \rho \right\| _{1,D}<1+\epsilon $.

\acknowledgements
A. O. Pittenger gratefully acknowledges the hospitality of the Centre for
Quantum Computation at Oxford University and support from UMBC and the
National Security Agency. M. H. Rubin wishes to thank the Office of Naval
Research and the National Security Agency for support of this work.

\section{\bf Appendix}

By emphasizing selected properties of projections for $d=2$, we can obtain a
representation of all (trace one) projections in spin notation for $d>2$. We
concentrate on $d=3$. To motivate the approach, recall from{\bf \ }(\ref
{2coef}{\bf )} that when $d=2$, $m_{3}=\left( 1\right) \rho _{0,0}+\left(
-1\right) \rho _{1,1}$, so that this particular spin coordinate is a convex
combination of $\left( +1\right) $ and $\left( -1\right) $, another way of
stating the well-known correspondence between $m_{3}$, the coefficient of $%
\sigma _{z}$, and the diagonal of $\rho $. If $\rho $ is also a projection,
then in the computational coordinates, $\rho _{j,k}=b_{j}b_{k}\exp \left(
i\left( \varphi _{j}-\varphi _{k}\right) \right) $, so that fixing $m_{3}$
fixes $\rho _{0,0}=b_{0}^{2}$ and $\rho _{1,1}=b_{1}^{2}$, and only the
phase factor $\theta =$ $\varphi _{0}-\varphi _{1}$ is unspecified. Using
the change of basis formula, the two remaining spin coefficients of a
projection with prescribed $m_{3}$ are thus given in terms of the parameter $%
\theta $ by 
\[
\left( 
\begin{array}{c}
m_{1} \\ 
-im_{2}
\end{array}
\right) =\left( 
\begin{array}{c}
s_{0,1} \\ 
s_{1,1}
\end{array}
\right) =\left( 
\begin{array}{cc}
1 & 1 \\ 
1 & -1
\end{array}
\right) \left( 
\begin{array}{c}
b_{0}b_{1}e^{i\theta } \\ 
b_{0}b_{1}e^{-i\theta }
\end{array}
\right) \text{,}
\]
where $0\leq \theta <2\pi $. If we let $t_{k}$ denote the value of $s_{k,1}$
when $\theta =0$, we can rewrite the preceding equation as 
\[
\left( 
\begin{array}{c}
s_{0,1} \\ 
s_{1,1}
\end{array}
\right) ==\frac{1}{2}\left( 
\begin{array}{cc}
1 & 1 \\ 
1 & -1
\end{array}
\right) \left( 
\begin{array}{cc}
e^{i\theta } & 0 \\ 
0 & e^{-i\theta }
\end{array}
\right) \left( 
\begin{array}{cc}
1 & 1 \\ 
1 & -1
\end{array}
\right) \left( 
\begin{array}{c}
t_{0} \\ 
t_{1}
\end{array}
\right) \text{.}
\]
Making the obvious definitions, this gives $\vec{s}=M_{2}\left( \theta
\right) \vec{t}$, and we also find that 
\[
M_{2}\left( \theta \right) =\left( 
\begin{array}{cc}
\cos \left( \theta \right)  & i\sin \left( \theta \right)  \\ 
i\sin \left( \theta \right)  & \cos \left( \theta \right) 
\end{array}
\right) =\cos \left( \theta \right) \sigma _{0}+i\sin \left( \theta \right)
\sigma _{x}.
\]
The geometry of this result is that if $-1<m_{3}<1$, then the remaining spin
coefficients in the projections associated with $m_{3}$\ can be identified
with the range of a one parameter family of invertible mappings $\left\{
M_{2}\left( \theta \right) \right\} $ acting on $\vec{t}$ and are
represented by the intersection of the surface of the Bloch sphere with a
horizontal plane at height $m_{3}$.

The same pattern of results holds for $d=3$. Since $s_{2,0}=s_{1,0}^{\ast }$%
, the diagonal of a given $\rho $ is in one-to-one correspondence with $%
s_{1,0}$ via the equation $s_{1,0}=\rho _{0,0}\left( 1\right) +\rho
_{1,1}\left( \eta ^{2}\right) +\rho _{2,2}\left( \eta \right) $. That is, $%
s_{1,0}$ is a convex combination of the vertices of an equilateral triangle
in the complex plane and thus {\it uniquely} corresponds to the weights of
the vertices, weights which are the entries of the diagonal of $\rho $. For
larger values of $d$, the geometry is more complicated. For example if $d=4,$
the diagonal of a given $\rho $ corresponds to two spin coefficients: $%
-1\leq s_{2,0}\leq +1$ and $s_{1,0}$ which is restricted to a rectangle in
the complex plane with vertices $\pm \left( 1+s_{2,0}\right) /2\pm i\left(
1-s_{2,0}\right) /2$. In general the diagonal of a density matrix $\rho $
corresponds to $d/2$ spin coefficients $s_{j,0}$, $j\neq 0$, when $d$ is
even and $\left( d-1\right) /2$ spin coefficients when $d$ is odd.

Once $s_{1,0}$ is fixed in the $d=3$ case, there are three complex
parameters remaining to be specified: $s_{0,1},s_{1,1},$ and $s_{2,1}$,
since the other four spin coefficients are forced by the restriction $%
s_{3-j,3-k}=\eta ^{jk}s_{j,k}^{\ast }$. If $\rho $ is a projection, $\sum
\left| s_{j,k}\right| ^{2}=3\sum \left| \rho _{j,k}\right| ^{2}=3$, and thus 
$\left| s_{2,0}\right| ^{2}+\sum \left| s_{k,1}\right| ^{2}=1$, tempting one
to look for an analogue of the Bloch sphere to represent all densities.
However, the normalization arising from $tr\left( \rho ^{2}\right) =1$ is
only a necessary condition on the parameters, and examples show it's not
sufficient. (See also \cite{arv}.) Instead we follow the $d=2$ paradigm and
describe trace one projections associated with a {\it fixed} $s_{1,0}$. If $%
\rho $ is such a projection, then in the computational coordinates $\rho
=\left| u\right\rangle \left\langle u\right| $, where $\left| u\right\rangle 
$ denotes a normalized three vector with $u_{k}=b_{k}e^{i\varphi _{k}}$ and $%
\sum \left| b_{k}\right| ^{2}=1$. Fixing $s_{1,0}$ fixes the $b_{k}$'s, and
it follows from Corollary 1 and the structure of $\rho $ that 
\[
\left( 
\begin{array}{c}
s_{0,1} \\ 
s_{1,1} \\ 
s_{2,1}
\end{array}
\right) =\left( 
\begin{array}{ccc}
1 & 1 & 1 \\ 
1 & \eta ^{2} & \eta \\ 
1 & \eta & \eta ^{2}
\end{array}
\right) \left( 
\begin{array}{ccc}
e^{i\theta _{0}} & 0 & 0 \\ 
0 & e^{i\theta _{1}} & 0 \\ 
0 & 0 & e^{i\theta _{2}}
\end{array}
\right) \left( 
\begin{array}{c}
b_{0}b_{1} \\ 
b_{1}b_{2} \\ 
b_{2}b_{0}
\end{array}
\right) 
\]
where $\theta _{k}=\varphi _{k}-\varphi _{k+1}$ with addition modulo $2\pi $
and with the normalization $\sum \theta _{k}=0 \bmod \,2\pi.$
Again letting $t_{k}$ denote the value of $s_{k,1}$ when the $\theta
_{k}$'s are chosen to be zero, this time we obtain a two parameter family of
projections associated with a given value of $s_{1,0}$. Letting $\vec{s}$
denote the column vector of parameters, $\vec{t}$ the column vector with
components $t_{k}$, and $\theta $ the $3-$vector of phase parameters, we
have $\vec{s}=M_{3}\left( \theta \right) \vec{t}$, where 
\begin{equation}
M_{3}\left( \theta \right) =\frac{1}{3}\left( 
\begin{array}{ccc}
1 & 1 & 1 \\ 
1 & \eta ^{2} & \eta \\ 
1 & \eta & \eta ^{2}
\end{array}
\right) \cdot \left( 
\begin{array}{ccc}
e^{i\theta _{0}} & 0 & 0 \\ 
0 & e^{i\theta _{1}} & 0 \\ 
0 & 0 & e^{i\theta _{2}}
\end{array}
\right) \cdot \left( 
\begin{array}{ccc}
1 & 1 & 1 \\ 
1 & \eta & \eta ^{2} \\ 
1 & \eta ^{2} & \eta
\end{array}
\right) =\sum_{k=0}^{2}f\left( k,\theta \right) S_{0,k}\text{.}
\label{parmat}
\end{equation}
If $\theta +\phi $ is defined as component-wise addition, then it is easy to
check that $\left\{ M_{3}\left( \theta \right) \right\} $ also defines an
Abelian group of invertible mappings, 
\[
M_{3}\left( \theta \right) \cdot M_{3}\left( \phi \right) =M_{3}\left(
\theta +\phi \right) \text{,} 
\]
giving the functional equation $\sum_{k}f\left( k,\theta \right) f\left(
j-k,\phi \right) =f\left( j,\theta +\phi \right) $ in analogy with the
corresponding result when $d=2.$ We have thus established a correspondence
between all trace one projection matrices with given diagonal and the range
of a two parameter family of mappings acting on $\vec{t}$. (We are endebted
to Rasmus Hansen for bringing to our attention \cite{bloore}, which contains
an analysis of the geometry of the convex space of $d=3$ densities. The
pre-tranform characterization of the projections associated with a given
diagonal is similar to the results derived here.)

In the $d=2$ case the choices of $m_{z}=\pm 1$ produce special cases of
projections, and the same is true when $d=3$. If $s_{1,0}$ is one of the
extreme points $1$, $\eta $, or $\eta ^{2}$, then two of the $b_{k}$'s equal
zero and all of the $s_{k,1}$'s equal zero. It follows that for $r=0$, $1$,
and $2$, $\frac{1}{3}\left[ S_{0,0}+\eta ^{r}S_{1,0}+\left( \eta
^{r}S_{1,0}\right) ^{\dagger }\right] $ is a trace one projection, and those
are the three subgroup projections $P_{1,0}(r).$ A degeneracy which has no
analogue in the $d=2$ case occurs when $s_{1,0}$ lies between two extreme
points on an edge. Then exactly one of the $b_{k}$'s equals zero, and there
is a one parameter family of projections associated with $s_{1,0}$. The most
interesting cases occur when $s_{1,0}$ lies in the interior of the
equilateral triangle. In particular when $s_{1,0}=0$, the $b_{k}$'s are
equal to $1/\sqrt{3}$, and by choosing the components of $\theta $
appropriately from $\left\{ 0,2\pi /3,4\pi /3\right\} $ we find the
remaining subgroup projections $P_{u}(r).$ Thus, our entire analysis of 
separability in the $d=3$ case uses only the projections associated with
the origin and with the vertices of the equilateral triangle.

\end{document}